\documentclass[showpacs,preprintnumbers,amsmath,amssymb,aps, prd]{revtex4}
\usepackage{graphicx}
\usepackage{grffile}
\usepackage{caption}
\usepackage{subcaption}
\usepackage{float}
\usepackage{dcolumn}
\usepackage{bm}
\begin{document}
\title{Bumblebee gravity with a Kerr-Sen-like solution and its Shadow}
\author{Sohan Kumar Jha}
\affiliation{Chandernagore College, Chandernagore, Hooghly, West
Bengal, India}
\author{Anisur Rahaman}
 \email{anisur.rahman@saha.ac.in;
 manisurn@gmail.com (Corresponding Author)}
\affiliation{Hooghly Mohsin College, Chinsurah, Hooghly - 712101,
West Bengal, India}

\date{\today}

\begin{abstract}
\begin{center}
Abstract
\end{center}
Lorentz-Violating (LV) scenario gets involved through a bumblebee
field vector field $B_\mu$. A spontaneous symmetry breaking allows
the field to acquires a vacuum expectation value that generates LV
into the system. A Kerr-Sen-like solution has been found out
starting from the generalized form of a radiating stationery
axially symmetric black-hole metric. We compute the effective
potential offered by the null geodesics in the bumblebee rotating
black-hole spacetime. The shadow has been sketched for different
variations of the parameters involved in the system. A careful
investigation has been carried out to study how the shadow gets
affected when Lorentz violation enters into the picture. The
emission rate of radiation has also been studied and how it varies
with the LV parameter $\ell$ is studied scrupulously.
\end{abstract}
\maketitle
\section{Introduction}
Various kinds of astronomical observations strongly confirm the
existence of the black-hole in the Universe. The recent message of
the detection of gravitational waves (GWs)\cite{GVW} by the $LIGO$
and $VIRGO$ observatories, and the captured image of the
black-hole shadow of a super-massive $M87^*$ black-hole by the
Event Horizon Telescope \cite{EVENT, EVENT1} provides substantial
evidence in support of the existence of black-hole. The perception
of the fundamental nature of spacetime would also likely to be
enriched with the information accessible from these recent
astronomical observations. So physics of black-hole acquire
renewed interest.

General relativity and the standard model of particle physics are
two very successful field theoretical models that assist us to
describe our Universe. The formulation of both these theories is
based on the well celebrated Lorentz symmetry. However, the regime
of applicability and the nature of service towards describing the
Universe by these two theories are profoundly different. The
general relativity describes the gravitational interaction and it
is a classical field theory in the curved spacetime and there is
no direct way to include the quantum effect. On the contrary, the
standard model describes the other fundamental interactions and it
is the quantum field theory in the flat spacetime that neglects
all gravitational effects, but to study the physical system in the
vicinity of the Plank scale $(10^{-19} GeV)$, the effect due to
gravity cannot be ignored, since the gravitational interaction is
strong enough in that energy scale. Therefore, the study of
physics in the vicinity of the Planck scale necessarily entails
the unification of general relativity and standard model particle
physics. Unfortunately, it is not yet developed with its full
wings because there is no straightforward way to quantize gravity.
Moreover, Lorentz invariance is not tenable in the regime where
spacetime is discrete in nature. So in the vicinity of the Planck
scale, it would not be unreasonable to discard Lorentz invariance.
On the other hand, Lorentz symmetry breaking in nature is
considered as an interesting and useful idea because it arises as
a possibility in the context of string theory \cite{STRING1,
STRING2, STRING3, STRING4}, noncommutative field theories
\cite{NONCOM} or loop quantum gravity theory \cite{LOOP1, LOOP2}.
As a consequence, nowadays Lorentz violation is considered as
relevant as well as a beneficial tool to probe the foundations of
modern physics. This fact suggests that signals associated with LV
is a promising way to investigate quantum gravity at the Planck
scale. The LV in the neutrino sector \cite{NEUTRINO}, the standard
model extension (SME) \cite{ESM}, and the LV effect on the
formation of atmospheric shower \cite{SHOAT} are some important
studies involving LV in this regard.

The effective theory that would  capture of the Plank scale effect
at least in a coarse-grained manner should encompass LV effect
through some LV sector. The Standard Model Extension (SME) is an
effective field theory which is worth mentioning that describes
the general relativity and the standard model at low energies,
that includes additional terms containing information about the LV
occurring at the Planck scale \cite{ESM, ESM1, ESM2, ESM3}. The
electromagnetic sector of the SME has been extensively studied in
the literature \cite{EMS1, EMS2, EMS3, EMS4, EMS5, EMS6, EMS7,
EMS8, EMS9, EMS10, EMS11, EMS12, EMS13, EMS14, EMS15, EMS16,
EMS17}. The electro-weak sector of it is described in the articles
\cite{EE1, EE2}. Furthermore, some effects of LV in the
gravitational sector have been studied in \cite{GVL1, GVL2, GVL3,
GVL4, GVL5}, specifically the case of the gravitational waves was
analyzed in the article \cite{GVW1, GVW2}.

For any theory of gravity, looking for a black-hole solution is a
significantly important extension, since black-holes provide into
the quantum gravity realm. In this respect, rotating black-hole
solutions are the most relevant subjects for astrophysics
\cite{KERR, KERRNU,ASEN}. It is known that the LV effect can be
instigated in the gravitational sector inviting the valuable
service of bumblebee field. In \cite{DING}, an attempt has been
made to find out an exact Kerr-like solution through solving
Einstein-bumblebee equations. In earlier work, Casana et al. found
an exact Schwarzschild-like solution in this bumblebee gravity
model and investigated its some classical tests \cite{RC}. Then
Rong-Jia Yang et al. study the accretion onto this black-hole
\cite{RONG} and find the LV parameter $\ell$ will slow down the
mass accretion rate. Sequential development entails that the
Kerr-sen-like solution from Einstein-bumblebee would also be of
worth investigation. It would be useful when the black-hole
contains both charge and angular momentum. We, therefore make an
attempt to seek a Kerr-sen-like solution from the
Einstein-bumblebee equations of motion.

The paper is organized as follows. Sect. II contains a review of
the Einstein-Bumblebee theory. In Sect. III, we find out an exact
Kerr-Sen-like black-hole solution from Einstein-Bumblebee theory
with the evaluation of Hawking temperature. Sect. IV is devoted to
the motion of photon and the sketching of black-hole shadow. How
this shadow deforms with $\ell$ is also investigated. In Sect. V,
the emission rate of radiation has been studied and how it varies
with $\ell$ is also observed closely. Sect. VI contains a summary
and discussion of the work.
\section{EINSTEIN-BUMBLEBEE THEORY}
Einstein-Bumblebee theory (or simply bumblebee model)is an
effective field theory where a vector field receives vacuum
expectation value (VEV)through spontaneous symmetry breaking. The
action of which is given by
\begin{eqnarray}
\mathcal{S}=\int d^{4} x \sqrt{-g}\left[\frac{1}{16 \pi
G_{N}}\left(\mathcal{R}+\varrho B^{\mu} B^{\nu} \mathcal{R}_{\mu
\nu}\right)-\frac{1}{4} B^{\mu \nu} B_{\mu
\nu}-V\left(B^{\mu}\right)\right]. \label{ACT}
\end{eqnarray}
where $\varrho^{2}$ is a real coupling constant (with mass
dimension -1) which controls the non-minimal gravity interaction
to bumblebee field $B_{\mu}$ (with the mass dimension 1). The
field strength tensor corresponding to the bumblebee field reads
\begin{eqnarray}
B_{\mu \nu}=\partial_{\mu} B_{\nu}-\partial_{\nu} B_{\mu}.
\end{eqnarray}
Through a spontaneous breaking of symmetry  a suitable potential
renders vacuum expectation value to the field $B_{\mu}$. The
conventional functional form of the potential
$V\left(B^{\mu}\right)$ that induces symmetry breaking is
\begin{eqnarray}
V=V\left(B_{\mu} B^{\mu}\pm b^2)\right.
\end{eqnarray}
$\mathrm {in}$ which $b^{2}$ is a real positive constant. It
donates a non-vanishing vacuum expectation value(VEV) for the
field $B_{\mu}$. This potential is assumed to have a minimum
through the condition
\begin{eqnarray}
B_{\mu}B^{\mu}\pm b^2=0. \label{MIN}
\end{eqnarray}
The Eqn. (\ref{MIN}) ensures that  a nonzero VEV,
$<B^{\mu}>=b^{\mu}$ will be supplied to the field $B_{\mu}$ by the
potential $V$. The vector $b^{\mu}$ is a function of the spacetime
coordinates having constant magnitude $b_{\mu} b^{\mu}=\mp b^{2}$.
Here $\pm$ signs indicate that $b^{\mu}$ may have a time-like as
well as for space-like nature depending upon the choice of sign.
The gravitational field equation in a vacuum that follows from the
action (\ref{ACT}) reads
\begin{eqnarray}
\mathcal{R}_{\mu \nu}-\frac{1}{2} g_{\mu \nu} \mathcal{R}=\kappa
T_{\mu \nu}^{B}. \label{GR}
\end{eqnarray}
where $\kappa=8 \pi G_{N}$ is the gravitational coupling, and
$T_{\mu \nu}^{B}$ is the bumblebee energy momentum tensor which is
given by the following expression:
\begin{eqnarray}\nonumber
T_{\mu \nu}^{B}&=&B_{\mu \alpha} B_{\nu}^{\alpha}-\frac{1}{4}
g_{\mu \nu} B^{\alpha \beta} B_{\alpha \beta}-g_{\mu \nu} V+2
B_{\mu} B_{\nu} V^{\prime} \\\nonumber
&+&\frac{\varrho}{\kappa}[\frac{1}{2} g_{\mu \nu} B^{\alpha}
B^{\beta} R_{\alpha \beta}
-B_{\mu} B^{\alpha} R_{\alpha \nu}-B_{\nu} B^{\alpha} R_{\alpha \mu} \\
&+&\frac{1}{2} \nabla_{\alpha} \nabla_{\mu}\left(B^{\alpha}
B_{\nu}\right)+\frac{1}{2} \nabla_{\alpha}
\nabla_{\nu}\left(B^{\alpha} B_{\mu}\right)-\frac{1}{2}
\nabla^{2}\left(B^{\mu} B_{\nu}\right)-\frac{1}{2} g_{\mu \nu}
\nabla_{\alpha} \nabla_{\beta} \left(B^{\alpha} B^{\beta}\right)].
\end{eqnarray}
Here prime(') denotes differentiation with respect to the
argument:
\begin{eqnarray}
V^{\prime}=\left.\frac{\partial V(x)}{\partial
x}\right|_{x=B^{\mu} B_{\mu} \pm b^{2}}.
\end{eqnarray}
Using the trace of Eqn. (\ref{GR}) we obtain the trace-reversed
version
\begin{eqnarray}
\mathcal{R}_{\mu \nu}=\kappa T_{\mu \nu}^{B}+2 \kappa g_{\mu \nu}
V-\kappa g_{\mu \nu} B^{\alpha} B_{\alpha}
V^{\prime}+\frac{\varrho}{4} g_{\mu \nu}
\nabla^{2}\left(B^{\alpha} B_{\alpha}\right)+\frac{\varrho}{2}
g_{\mu \nu} \nabla_{\alpha} \nabla_{\beta}\left(B^{\alpha}
B^{\beta}\right).
\end{eqnarray}
One immediately sees that when the bumblebee field $B_{\mu}$
vanishes, we recover the ordinary Einstein equations.The equation
of motion for the bumblebee field is
\begin{eqnarray}
\nabla^{\mu} B_{\mu \nu}=2 V^{\prime}
B_{\nu}-\frac{\varrho}{\kappa} B^{\mu} R_{\mu \nu}.
\end{eqnarray}
When the bumblebee field remains frozen in its vacuum expectation
value we are allowed to write
\begin{eqnarray}\nonumber
B_{\mu}&=&b_{\mu}\\
b_{\mu \nu}&=&\partial_{\mu} b_{\nu}-\partial_{\nu} b_{\mu}.
\end{eqnarray}
Under this condition form of the potential is irrelevant which is
important is
\begin{eqnarray}
V=0, \qquad \qquad V^{\prime}=0.
\end{eqnarray}
In this situation, Einstein equations acquires a generalized form:
\begin{eqnarray}\nonumber
&&\mathcal{R}_{\mu \nu}-\kappa b_{\mu \alpha}
b_{\nu}^{\alpha}+\frac{\kappa}{4} g_{\mu \nu} b^{\alpha \beta}
b_{\alpha \beta}+\varrho b_{\mu} b^{\alpha} \mathcal{R}_{\alpha
\nu}+\varrho b_{\nu} b^{\alpha} \mathcal{R}_{\alpha
\mu}\\\nonumber &&-\frac{\varrho}{2} g_{\mu \nu} b^{\alpha}
b^{\beta} \mathcal{R}_{\alpha \beta}
-\frac{\varrho}{2}\left[\nabla_{\alpha}
\nabla_{\mu}\left(b^{\alpha} b_{\nu}\right) +\nabla_{\alpha}
\nabla_{\nu}\left(b^{\alpha} b_{\mu}\right)
-\nabla^{2}\left(b_{\mu} b_{\nu}\right)\right]=0\\
&&\Rightarrow\bar{R}_{\mu \nu}=0,\label{MEQ}
\end{eqnarray}
with
\begin{eqnarray}\nonumber
\bar{R}_{\mu \nu}&=&\mathcal{R}_{\mu \nu}-\kappa b_{\mu \alpha}
b_{\nu}^{\alpha} +\frac{\kappa}{4} g_{\mu \nu} b^{\alpha \beta}
b_{\alpha \beta} +\varrho b_{\mu} b^{\alpha} \mathcal{R}_{\alpha
\nu} +\varrho b_{\nu} b^{\alpha} \mathcal{R}_{\alpha \mu}
-\frac{\varrho}{2} g_{\mu \nu} b^{\alpha} b^{\beta}
\mathcal{R}_{\alpha \beta}
+\bar{B}_{\mu \nu} \\
\bar{B}_{\mu \nu}&=&-\frac{\varrho}{2}\left[\nabla_{\alpha}
\nabla_{\mu}\left(b^{\alpha} b_{\nu}\right)+\nabla_{\alpha}
\nabla_{\nu}\left(b^{\alpha}
b_{\mu}\right)-\nabla^{2}\left(b_{\mu} b_{\nu}\right)\right].
\end{eqnarray}
With this input we  proceed to find out the exact Kerr-Sen-like
solution from the Einstein-bumblebee model.
\section{EXACT KERR-SEN-LIKE SOLUTION IN EINSTEIN-BUMBLEBEE MODEL}
An attempt is made here to to find out the exact Kerr-Sen-like
solution from the Einstein-bumblebee model. We will follow the
similar guideline as Koltz to reproduce the Kerr solution
\cite{KOLTZ, KOLTG} did to offer the exact Kerr solution. In the
article \cite{DING}, a Kerr-like solution has been extracted out
following the same development of Koltz to reproduce the Kerr
solution \cite{KOLTZ, KOLTG}. According to the development of
Koltz, the generalized form of radiating stationary axially
symmetric black-hole metric can be written down as \cite{KOLTZ,
KOLTG, DING}
\begin{eqnarray}
d s^{2}=-\gamma(\zeta, \theta) d
\tau^{2}+a[p(\zeta)-q(\theta)]\left(d \zeta^{2}+d
\theta^{2}+\frac{q}{a} d \phi^{2}\right)-2q(\theta)d\tau d\phi.
\label{METRIC}
\end{eqnarray}
where $a$ is a dimensional constant which is introduced to take
care of dimensional agreement. The time $t$  and $\tau$ has the
relation
\begin{eqnarray}
d \tau=d t-q d \phi.
\end{eqnarray}
In terms of $t$  Eqn. (\ref{METRIC}) turns into
\begin{eqnarray}\nonumber
&d s^{2}&=-\gamma(\zeta, \theta) d t^{2}+a[p(\zeta)
-q(\theta)]\left(d \zeta^{2}+d \theta^{2}\right) \\
&+&\left\{[1-\gamma(\zeta, \theta)] q^{2}(\theta)+p(\zeta)
q(\theta)\right\} d \phi^{2}-2 q(\theta)[1-\gamma(\zeta, \theta)]
d t d \phi.\label{METRIC1}
\end{eqnarray}
We now use this metric ansatz (\ref{METRIC1}) to compute the
gravitational field equations. If we consider that bumblebee field
is space-like it can be casted in the form
\begin{eqnarray}
b_{\mu}=(0, b(\zeta, \theta), 0,0).
\end{eqnarray}
Our focus will be laid on the bumblebee field that will acquire a
pure radial VEV. It is reasonable to consider the space-like
nature of the bumblebee field, since in this situation, space-time
curvature has greater radial variation compared to its temporal
variation. Now we have
\begin{eqnarray}\nonumber
& &b_{\mu} b^{\mu}=b_{0}^{2},\\\nonumber &\Rightarrow& g^{\mu \nu}
b_{\mu} b_{\nu}=b_{0}^{2},\\\nonumber
&\Rightarrow& \frac{b^{2}}{a\left(p-q\right)}=b_{0}^{2},\\
&\Rightarrow&  b=b_{0} \sqrt{a\left(p-q\right)}.
\end{eqnarray}
where $b_{0}$ is a constant. Hence the explicit form of $b_{\mu}$
comes out to
\begin{eqnarray}
b_{\mu}=\left(0, b_{0} \sqrt{a(p-q)}, 0,0\right),\label{FIELD}
\end{eqnarray}
in a straightforward manner. Therefore, the non-vanishing
components of the bumblebee field are
\begin{eqnarray}\nonumber
 b_{\zeta \theta}&=&-b_{\theta \zeta}\\\nonumber
&=&\partial_{\zeta}b_{\theta}-\partial_{\theta}b_{\zeta}\\\nonumber
&=&-\frac{\partial}{\partial \theta}b\left(\zeta,\theta\right)=
-\frac{\partial}{\partial \theta}b_{0}\sqrt{a\left(p-q\right)}\\
&=&\frac{a b_{0} q^{\prime}}{ 2 \sqrt{a(p-q)}}.
\end{eqnarray}
where the prime is used to indicate a derivative with respect to
its argument. In addition the quantity $b_{\mu}^{\alpha} b_{\nu
\alpha}$ has the following non-vanishing components:
\begin{eqnarray}\nonumber
b_{\zeta}^{\alpha} b_{\zeta \alpha}&=&b_{\theta}^{\alpha}
b_{\theta \alpha}\\\nonumber &=&g^{\alpha \beta}b_{\theta
\beta}b_{\theta \alpha}
=g^{\zeta \zeta}b_{\theta \zeta}b_{\theta \zeta}\\
&=&\frac{b_{0}^{2} q^{\prime 2}}{4(p-q)^{2}},
\end{eqnarray}
and the quantity $b^{\alpha \beta} b_{\alpha \beta}$ has a
non-vanishing contribution
\begin{eqnarray}\nonumber
b^{\alpha \beta} b_{\alpha \beta}&=&g^{\mu \alpha}g^{\nu
\beta}b_{\mu \nu}b_{\alpha \beta}\\\nonumber &=&g^{\mu
\zeta}g^{\nu \theta}b_{\mu \nu}b_{\zeta \theta}
+g^{\mu \theta}g^{\nu \zeta}b_{\mu \nu}b_{\theta \zeta}\\
&=&\frac{b_{0}^{2} q^{\prime 2}}{2 a(p-q)^{3}}.
\end{eqnarray}
For the metric (\ref{METRIC1}) the nonzero components of Ricci
tensor are $\mathcal{R}_{t t}, \mathcal{R}_{t \phi},
\mathcal{R}_{\zeta \zeta}, \mathcal{R}_{\zeta \theta},
\mathcal{R}_{\theta \theta}, \mathcal{R}_{\phi \phi}$. It is
straightforward to see that $\bar{B}_{\zeta \theta}=0$. The
gravitational field equations which are needed for our purpose are
\begin{eqnarray}\nonumber
\bar{R}_{\zeta \theta}&=&\mathcal{R}_{\zeta \theta}-\kappa
b_{\zeta \alpha}b^{\alpha}_{\theta}+\varrho
b_{\zeta}b^{\alpha}\mathcal{R}_{\alpha \theta}+\varrho
b_{\theta}b^{\alpha}\mathcal{R}_{\alpha \mu}\\\nonumber
&=&\mathcal{R}_{\zeta \theta}-\kappa b_{\zeta
\alpha}b^{\alpha}_{\theta}+\varrho b^{2}_{0}\mathcal{R}_{\zeta
\theta}\\\nonumber &=&(1+\ell) \mathcal{R}_{\zeta \theta}
\\\nonumber
\bar{R}_{t t}&=&\mathcal{R}_{t t}+g_{t t}\left(\frac{\kappa}{4}
b^{\alpha \beta} b_{\alpha \beta}
-\frac{\varrho}{2} b^{\zeta} b^{\zeta} \mathcal{R}_{\zeta \zeta}\right)+\bar{B}_{t t} \\
\bar{R}_{t \phi}&=&\mathcal{R}_{t \phi}+g_{t
\phi}\left(\frac{\kappa}{4} b^{\alpha \beta} b_{\alpha
\beta}-\frac{\varrho}{2} b^{\zeta} b^{\zeta} \mathcal{R}_{\zeta
\zeta}\right)+\bar{B}_{t \phi} \label{RBAR}
\end{eqnarray}
where $\ell=\varrho b_{0}^{2}$. The quantities $\mathcal{R}_{\zeta
\theta}, \bar{B}_{t t}, \bar{B}_{t \phi}$ are as
\begin{eqnarray}\nonumber
&\mathcal{R}_{\zeta \theta}&=-\frac{\bar{\Delta}_{12}}{2
\bar{\Delta}}+\frac{\Delta_{2}\left[(p-q) \bar{\Delta}_{1}+2
\bar{\Delta} p_{1}\right]}{4(p-q) \bar{\Delta}^{2}} \\\nonumber
&\bar{B}_{t t}&=\ell\left[\frac{\gamma_{11}}{2 a(p-q)}
+\frac{\gamma}{4 a(p-q) \bar{\Delta}} p_{1} \gamma_{1}
-\frac{1}{4 a \bar{\Delta}} \gamma_{1}^{2}\right] \\
&\bar{B}_{t \phi}&=\ell\left[-\frac{q \gamma_{11}}{2
a(p-q)}+\frac{q(2-\gamma)}{4 a(p-q) \Delta} p_{1}
\gamma_{1}+\frac{q}{4 a \bar{\Delta}} \gamma_{1}^{2}\right].
\label{RB BAR}
\end{eqnarray}
where $\bar{\Delta}=q+\gamma(p-q)$.  The differentiation  with
respect to the variable $\zeta$ and $\theta$ are denoted by the
suffixes $1$ and $2$ respectively in the Eqn. (\ref{RB BAR}). Note
that $\bar{R}_{\zeta \theta}=0$ implies that $\mathcal{R}_{\zeta
\theta}$ vanishes. This helps us to set  $\bar{\Delta}_{2}=0$
which in turn yields
\begin{eqnarray}\nonumber
&&\gamma_{2}p+\left(1-\gamma\right)q_{2}-\gamma_{2}q=0,\\
&&\Rightarrow \gamma_{2}=-\frac{(1-\gamma) q_{2}}{p-q}.
\end{eqnarray}
A few steps of algebra gives the expression  $\gamma$:
\begin{eqnarray}
\gamma=1-\frac{2 h(\zeta)}{p(\zeta)-q(\theta)}.
\end{eqnarray}
We are allowed to introduce new independent variable $\sigma$
exploiting the condition  $\bar{\Delta}_{2}=0$:
\begin{eqnarray}
\sigma=\int \sqrt{\bar{\Delta}} d \zeta,
\end{eqnarray}
where $\bar{\Delta}=p-2 h$.  Now taking  the derivatives f $p$
with respect to $\zeta$ we have
 \begin{eqnarray}
p_{1}&=&\frac{d p}{d \zeta}=\frac{d \sigma}{d \zeta} \frac{d p}{d
\sigma}=\sqrt{\bar{\Delta}} \frac{d p}{d \sigma},\\\nonumber
p_{11}&=&\frac{d^{2} p}{d \zeta^{2}}=\frac{d}{d
\zeta}\left(\sqrt{\bar{\Delta}}\frac{dp}{d
\sigma}\right),\\\nonumber &=&\sqrt{\bar{\Delta}}\frac{d^{2}p}{d
\sigma^{2}}\frac{d \sigma}{d \zeta}+\frac{dp}{d \sigma}\frac{d}{d
\zeta}\left(\sqrt{\bar{\Delta}}\right),\\\nonumber &=&\bar{\Delta}
\frac{d^{2} p}{d \sigma^{2}}+\frac{1}{2}\left(\frac{d p}{d
\sigma}\right)^{2}-\frac{d h}{d \sigma} \frac{d p}{d \sigma}.
\end{eqnarray}
A careful look on the Eqns. (\ref{MEQ}) and (\ref{RBAR}), reveals
that the following equation holds:
\begin{eqnarray}
g_{t \phi} \bar{R}_{t t}-g_{t t} \bar{R}_{t \phi}=0. \label{COMB}
\end{eqnarray}
After inserting the expressions of  $g_{t \phi}$, $\bar{R}_{t t}$,
$g_{t t}$ and $\bar{R}_{t \phi}$  in (\ref{COMB}), we find
\begin{eqnarray}\nonumber
&p&\left[4(1+\ell) \frac{\dot{h} \dot{p}}{h} q^{2}-2 q
q_{22}+q_{2}^{2}+2(1+\ell) \ddot{p} q^{2}\right] \\\nonumber
&-&4(1+\ell) \dot{p}^{2} q^{2}-2(p-q)^{2} q^{2} \ddot{h}\left(1+\frac{\ell}{h}\right) \\
&-&q\left[4(1+\ell) \frac{\dot{h} \dot{p}}{h} q^{2}-2 q q_{22}+5
q_{2}^{2}+2(1+\ell) \ddot{p} q^{2}\right]=0, \label{COMBINED}
\end{eqnarray}
where dot denotes derivative with respect to $\sigma$.  Here $p$
and $h$ are functions of $\sigma$,  and $q$ is a function of
$\theta$. We, therefore, come up to
\begin{eqnarray}
\frac{\dot{h} \dot{p}}{h}=k, \dot{p}^{2}=c p+n, \ddot{h}=0,
\label{CONDITION}
\end{eqnarray}
without any loss of generality.  Note that $k$, $c$, $n$ are some
constants. We find that $\ddot{p}=k= \frac{c}{2}$ and Eqn.
(\ref{COMBINED}) reduces  to the following
\begin{eqnarray}\nonumber
4(1+\ell)(k-c) q^{2}-2 q q_{22}+q_{2}^{2}+(1+\ell) c q^{2}=0,\\
4k(1+\ell) q^{2}-2 q q_{22}+5 q_{2}^{2}+(1+\ell) c q^{2}+4(1+\ell)
n q=0. \label{DEG}
\end{eqnarray}
They both the equations in (\ref{DEG}) givie
\begin{eqnarray}
q_{2}^{2}=-(1+\ell)\left(c q^{2}+n q\right).
\end{eqnarray}
So we come up to
\begin{eqnarray}
q=-\frac{n}{c} \sin ^{2}[\sqrt{(1+\ell) c} \frac{\theta}{2}].
\end{eqnarray}
By setting the constants $c=4 /(1+\ell)$ and $n=-4a$ it becomes
\begin{eqnarray}
q=(1+\ell) a \sin ^{2} \theta
\end{eqnarray}
From the conditions (\ref{CONDITION}), we find that
\begin{eqnarray}
p=\frac{\sigma^{2}}{1+\ell}+a(1+\ell), h=c^{\prime} \sigma,
\gamma=1-\frac{2(1+\ell) c^{\prime}
\sigma}{\sigma^{2}+a\left(1+\ell \right)^{2} \cos ^{2} \theta},
\end{eqnarray}
where $c^{\prime}$ is a constant. After choosing
$\sigma=\sqrt{\frac{\ell+1}{a}\frac{r+b}{r}} r, c^{\prime}=M /
\sqrt{(\ell+1) a\frac{r+b}{r}}$ and $\phi=\varphi / \sqrt{1+\ell}$
for Boyer-Lindquist coordinates, we  arrive at
\begin{eqnarray}
p=\frac{r(r+b)}{a}+a(\ell+1), h=\frac{M r}{a}, \gamma=1-\frac{2 M
r}{\rho^{2}},
\end{eqnarray}
 where $\rho^{2}=r(r+b)+(1+\ell) a^{2} \cos ^{2} \theta$.
 Finally, substituting these quantities
into the Eqns. (\ref{METRIC1}) and (\ref{FIELD}), we obtain the
bumblebee field $b_{\mu}=\left(0, b_{0} \rho, 0,0\right)$ and the
rotating metric in the bumblebee gravity is found out to
\begin{eqnarray}
d s^{2}=-\left(1-\frac{2 M r}{\rho^{2}}\right) d t^{2}-\frac{4 M r
a \sqrt{1+\ell} \sin ^{2} \theta}{\rho^{2}} d t d
\varphi+\frac{\rho^{2}}{\Delta} d r^{2}+\rho^{2} d
\theta^{2}+\frac{A \sin ^{2} \theta}{\rho^{2}} d \varphi^{2},
\label{FINAL}
\end{eqnarray}
where
\begin{eqnarray}
\Delta=\frac{r(r+b)-2 M r}{1+\ell}+a^{2}, A=\left[r(r+b)+(1+\ell)
a^{2}\right]^{2}-\Delta(1+\ell)^{2} a^{2} \sin ^{2} \theta.
\end{eqnarray}
If $\ell \rightarrow 0$ it recovers the usual Kerr-Sen  metric and
for $a \rightarrow 0$, and $b \rightarrow 0$,  it becomes
\begin{eqnarray}
d s^{2}=-\left(1-\frac{2 M}{r}\right) d t^{2}+\frac{1+\ell}{1-2 M
/ r} d r^{2}+r^{2} d \theta^{2}+r^{2} \sin ^{2} \theta d
\varphi^{2},
\end{eqnarray}
which is identical to the metric obtained in the article
\cite{RC}. The metric (\ref{FINAL}), represents a purely radial
Lorentz-violating black-hole solution with rotating angular
momentum $J=\frac{a}{M}$ and charge $Q=\sqrt{bM}$. It is singular
at $\rho^{2}=0$ and at $\Delta=0$. Its event horizons and
ergosphere are located  at
\begin{eqnarray}
r_{\pm}=M-\frac{b}{2} \pm\frac{
\sqrt{(b-2M)^{2}-4a^{2}(1+\ell)}}{2}, r_{\pm}^{e r g o}=M
-\frac{b}{2}\pm \frac{\sqrt{(b-2M)^{2}-4a^{2}(1+\ell) \cos ^{2}
\theta}}{2},
\end{eqnarray}
where $\pm$ signs correspond to the outer and inner
horizon/ergosphere respectively. So there exists a black-hole if
and only if
\begin{eqnarray}
|b-2M| \geq 2|a|\sqrt{1+\ell}.
\end{eqnarray}
Let us now calculate the Hawking temperature corresponding to this
black-hole which we obtained from its surface gravity \cite{RM} as
follows
\begin{eqnarray}
T=\frac{\kappa}{2 \pi}, \kappa=-\frac{1}{2} \lim _{r \rightarrow
r_{+}} \sqrt{\frac{-1}{X}} \frac{d X}{d r}, X \equiv g_{t
t}-\frac{g_{t \varphi}^{2}}{g_{\varphi \varphi}}.
\end{eqnarray}
Inserting corresponding metric components in Eqn. (\ref{FINAL}),
we get
\begin{eqnarray}
T=\frac{\sqrt{\left(2M-b\right)^2-4a^2\left(1+l\right)}} {4\pi M
\sqrt{1+l}\left(2M-b+\sqrt{\left(2M-b\right)^2
-4a^2\left(1+l\right)}\right)}.
\end{eqnarray}
\section{PHOTON ORBIT AND BLACK-HOLE SHADOW}
A black-hole shadow is an optical appearance that occurs when
there is a bright distant light source behind the black-hole.
Although it is not established clearly, it would be reasonable to
believe that the proper investigation on black-hole shadow  will
provides considerable insight about nature of black-hole. It looks
like a two-dimensional dark zone for a distant observer. Synge in
\cite{SYNGE} studied the shadow of the Schwarzschild black-hole.
He pointed out that the edge of the shadow is rounded. Bardeen in
the article \cite{BARDEEN} studied the shadow of the Kerr
black-hole and he argued that this shadow is no longer circular.
The articles \cite{KH, SDO1, SDO2, SDO3, SDO4, SDO5, SDO6, TT,
DASTAN} contain more recent research on this issue. We are
intended to study the LV effect on the black-hole shadow. In this
context we have considered the Kerr-Sen Black-hole with a
bumblebee background and study how the shadow of this black-hole
gets affected (deformed) by the Lorentz-violating parameter
associated with the bumblebee field.

In order to study black-hole shadow we introduce two conserved
parameters $\xi$ and $\eta$ which are defined by
\begin{eqnarray}
\xi=\frac{L_{z}}{E}, \eta=\frac{\mathcal{Q}}{E^{2}},
\end{eqnarray}
respectively, where $E, L_{z}$, and $\mathcal{Q}$ are the energy,
the axial component of the angular momentum, and Carter constant,
respectively. Then the null geodesics in the bumblebee rotating
black-hole spacetime in terms of $\xi$ are given by
\begin{eqnarray}\nonumber
\rho^{2} \frac{d r}{d \lambda}=\pm \sqrt{R}, \rho^{2} \frac{d
\theta}{d \lambda}=\pm \sqrt{\Theta}, \\\nonumber
(1+\ell) \Delta \rho^{2} \frac{d t}{d \lambda}=A-2 \sqrt{1+\ell} \operatorname{Mra\xi}, \\
(1+\ell) \Delta \rho^{2} \frac{d \phi}{d \lambda}=2 \sqrt{1+\ell}
M r a+\frac{\xi}{\sin ^{2} \theta}\left(\rho^{2}-2 M r\right),
\end{eqnarray}
where $\lambda$ is the affine parameter, and
\begin{eqnarray}
R(r)=\left[\frac{r(r+b)+(1+\ell) a^{2}}{\sqrt{1+\ell}}-a
\xi\right]^{2}-\Delta\left[\eta+(\xi-\sqrt{1+\ell}a)^{2}\right],
\Theta(\theta)=\eta+(1+\ell) a^{2} \cos ^{2} \theta-\xi^{2} \cot
^{2} \theta.
\end{eqnarray}

\begin{figure}[H]
\centering
\begin{subfigure}{.5\textwidth}
\centering
\includegraphics[width=.7\linewidth]{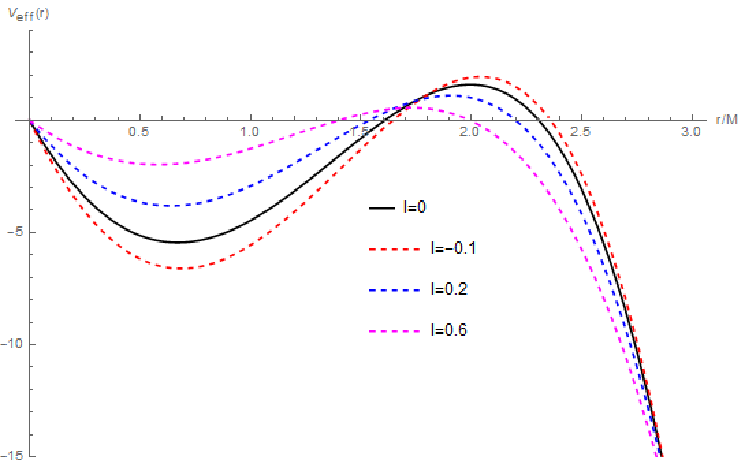}
\end{subfigure}%
\begin{subfigure}{.5\textwidth}
 \centering
  \includegraphics[width=.7\linewidth]{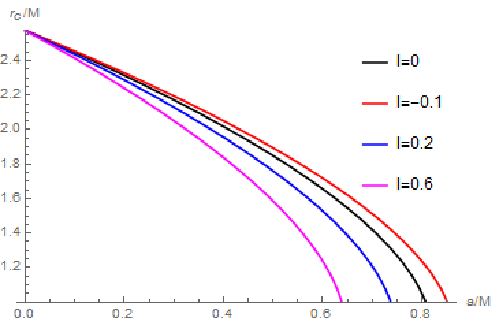}
\end{subfigure}
\caption{The left panel describes the effective potential for
various values of $l$ with $b/M=.36,a/M=.5$ and right panel
describes critical radius for various values of $l$ with $b=.36$ }
\begin{subfigure}{.5\textwidth}
  \centering
  \includegraphics[width=.7\linewidth]{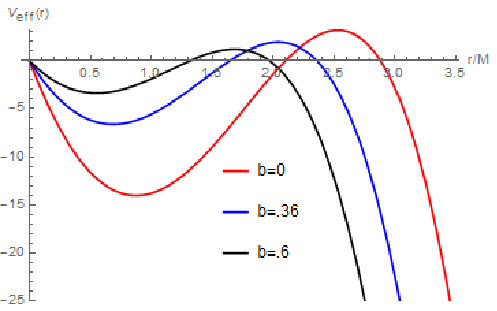}
\end{subfigure}%
\begin{subfigure}{.5\textwidth}
  \centering
  \includegraphics[width=.7\linewidth]{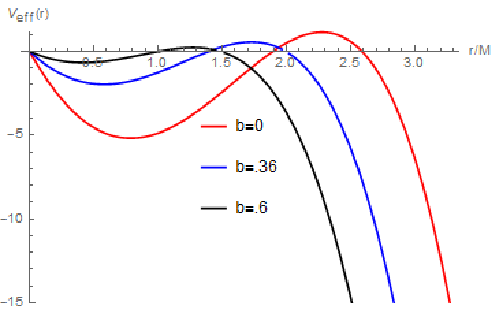}
\end{subfigure}
\caption{The left panel describes effective potential for various
 values of $b$ with $a=.5,l=-.1$, and right panel describes effective
  potential for various values of $b$ with $a=.5,l=.6$}
\end{figure}

\begin{figure}[H]
\centering
\begin{subfigure}{.5\textwidth}
  \centering
  \includegraphics[width=.7\linewidth]{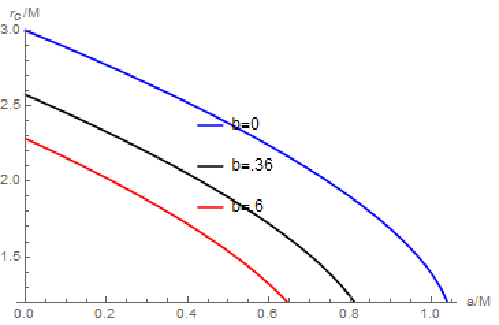}
\end{subfigure}%
\begin{subfigure}{.5\textwidth}
  \centering
  \includegraphics[width=.7\linewidth]{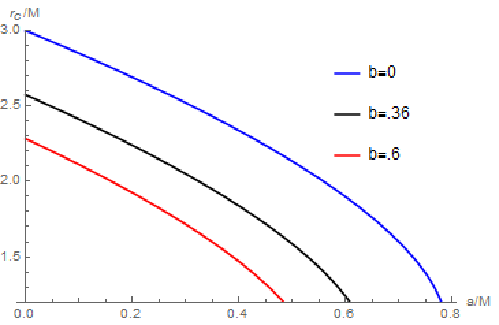}
\end{subfigure}
\caption{The left panel describes critical radius for various
values of $b$ with $l=-.1$, and right panel describes critical
radius for various values of $b$ with $l=.6$  }
\end{figure}

The radial equation of motion can be written down in the form
\begin{eqnarray}
\left(\rho^{2} \frac{d r}{d \lambda}\right)^{2}+V_{e f f}=0,
\end{eqnarray}
where the effective potential $V_{e f f}$ reads
\begin{eqnarray}
V_{e f f}=-[\frac{r(r+b)+(1+\ell) a^{2}}{\sqrt{1+\ell}}-a
\xi]^{2}+\Delta\left[\eta+(\xi-\sqrt{1+\ell}a)^{2}\right].
\end{eqnarray}
Note that   $V_{e f f}(0)=0$, and $\quad V_{e f f}(r \rightarrow
\infty) \rightarrow \infty $.
The unstable spherical orbit on the
equatorial plane is given by the following equations
\begin{eqnarray}
\theta=\frac{\pi}{2}, R(r)=0, \frac{d R}{d r}=0, \frac{d^{2} R}{d
r^{2}}<0, \eta=0.\label{CONDITION1}
\end{eqnarray}
We plot the $V_{e f f}$ against $r/M$ with $\xi=\xi_{c}+0.2$ where
$\xi_{c}$ is the value of $\xi$ for equatorial spherical unstable
direct orbit. Plots show that the photon starting from infinity
will meet the turning point, and then turns back to infinity. When
$\xi=\xi_{c},$ this turning point is an unstable spherical orbit
which gives the boundary of the shadow \cite{SW}. It also shows
that the deviation from GR (Kerr-Sen): when $\mathrm{LV}$ constant
$\ell > 0$, the turning point shifts to the left however when
$\ell < 0$ it shifts to the right. These shifts are similar to
those of the Einstein-aether black-hole \cite{TT}, which is also
the LV black-hole. The same type of shifting was reported in
\cite{DING} for Kerr-like black-hole. We have also plotted radius
$r_{c}$ of unstable equatorial spherical direct orbit against
$a/M$ for various scenarios. It shows that the $r_{c}$ decreases
with $\ell
>0 $, however it increases when $\ell <0$, which are similar to
those of the noncommutative black-hole \cite{SW}. This is also
similar to the observation of the article \cite{DING}. The $r_{c}$
decreases with $b$ for a particular $\ell$ irrespective of its
sign.

For more generic orbits $\theta \neq \pi / 2$ and $\eta \neq 0,$
the solution of Eqn. (\ref{CONDITION1})$ r=r_{s},$ gives the $r-$
constant orbit, which is also called spherical orbit and the
conserved parameters of the orbits are given by
\begin{equation}
\begin{aligned}
\xi_{s} &=\frac{a^{2}\left(1+\mathrm{l}\right)\left(2 M+2 r_{\mathrm{s}}
+b\right)+r_{\mathrm{s}}\left(2 r_{\mathrm{s}}^{2}
+3 b r_{\mathrm{s}}+b^{2}-2 M\left(3 r_{\mathrm{s}}
+b\right)\right)}{a\sqrt{1+\mathrm{l}}\left(2 M-2 r_{\mathrm{s}}-b\right)} \\
\eta_{s} &=-\frac{r_{\mathrm{s}}^{2}\left(-8 a^{2}\left(1+\mathrm{l}\right)
M\left(2 r_{\mathrm{s}}+b\right)
+\left(2 r_{\mathrm{s}}^{2}+3b r_{\mathrm{s}}
+b^{2}
-2 M\left(3 r_{\mathrm{s}}+b\right)\right)^{2}\right)}{a^{2}\left(1+\mathrm{l}\right)
\left(2 M-2 r_{\mathrm{s}}-b\right)^{2}} .
\end{aligned}
\end{equation}
The two celestial coordinates, which are used to describe the
shape of the shadow that an observers see in the sky, can be given
by
\begin{eqnarray}\nonumber
\alpha(\xi, \eta ; \theta)&=&\lim _{r \rightarrow \infty} \frac{-r
p^{(\varphi)}}{p^{(t)}} = -\xi_{s} \csc \theta,\\\nonumber
\beta(\xi, \eta ; \theta)&=&\lim _{r \rightarrow \infty} \frac{r
p^{(\theta)}}{p^{(t)}} =\sqrt{\left(\eta_{s}+a^{2} \cos ^{2}
\theta-\xi_{s}^{2} \cot ^{2} \theta\right)}
\end{eqnarray}
where $\left(p^{(t)}, p^{(r)}, p^{(\theta)}, p^{(\phi)}\right)$
are the tetrad components of the photon momentum with respect to
locally non-rotating reference frames \cite{BARDEEN}.

The shadow of the collapsed object is defined as follows. Suppose
some light rays are emitted at infinity ($r = +\infty$) and
propagate near the collapsed object. If these rays can reach the
observer at infinity after scattering, then that direction would
not appear as dark. On the other hand, when they will be incident
into the event horizon of a black-hole, the observer will never
see such light rays. Such a direction will look dark and that
ultimately appears as a shadow. We define the apparent shape of a
black-hole by the boundary of the shadow \cite{PJ}. We have
studied the shapes of the shadow for different variations of the
parameters involved in the theory with a special emphasis on the
variation of $\ell$, With the increase of the $\mathrm{LV}$
parameter $\ell$, its left endpoint moves to the right and then
the right endpoint though moves towards right, the movement is
very little compared to the movement of the left end so a deformed
shape is viewed.
\begin{figure}[H]
\centering
\begin{subfigure}{.5\textwidth}
  \centering
  \includegraphics[width=.7\linewidth]{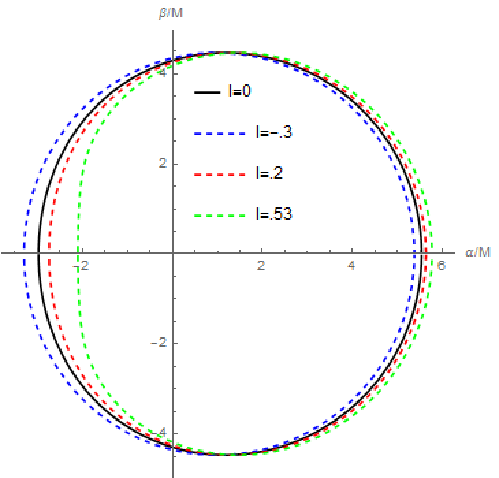}
\end{subfigure}%
\begin{subfigure}{.5\textwidth}
\centering
  \includegraphics[width=.7\linewidth]{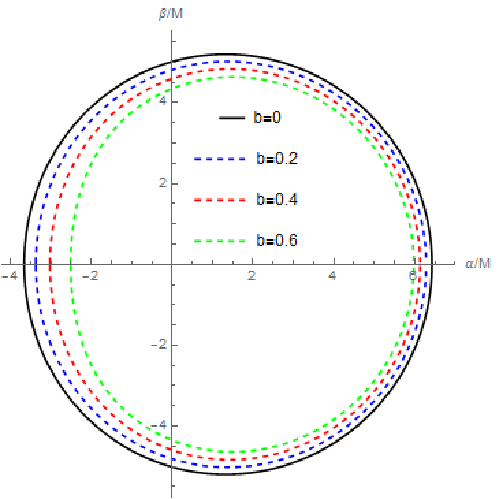}
\end{subfigure}%
\caption{The left one gives shapes of the shadow for various
values of $l$ with $a/M=0.5$, $b=0.76$,
 and $\theta=\pi/2$, and the right one gives shapes of the shadow for
  various values of $b$ with $a/M=0.7$ , $l=-.1$, and $\theta=\pi/2$}
\end{figure}

\begin{figure}[H]
\centering
\begin{subfigure}{.5\textwidth}
  \centering
  \includegraphics[width=.7\linewidth]{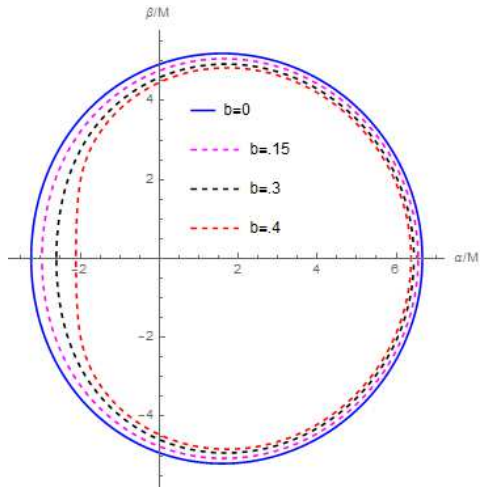}
\end{subfigure}
\caption{ The shapes of the shadow for various values of $b$
with $a/M=0.7$, $l=0.3$ and $\theta=\pi/2$ }
\end{figure}
The Fig. $4$, shows that the left end of the shadow shifts towards
the right for positive values of $\ell$ for a particular $b$ and
the reverse is the case when $\ell$ is negative. However
irrespective of the sign of $\ell$ the shifting of the left end of
the shadow is towards the right for increasing $b$, See Fig. $5$.
Using the parameters which are introduced by Hioki and Maeda
\cite{KH}, we now turn to analyze deviation from circular form
$\left(\delta_{s}\right)$ and the size $\left(R_{s}\right)$ of the
shadow image of the black-hole as follows.
\begin{figure}[H]
\centering
\begin{subfigure}{.5\textwidth}
  \centering
  \includegraphics[width=.7\linewidth]{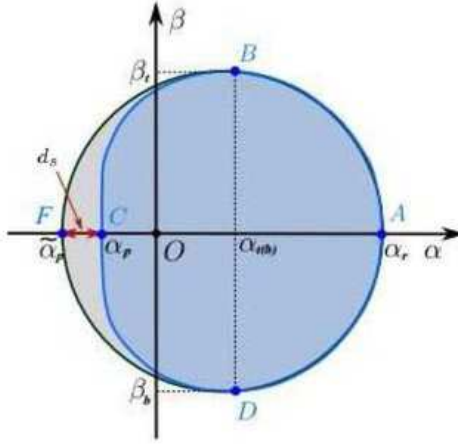}
\end{subfigure}
\caption{The black-hole shadow and reference circle. $d_s$ is the
distance between the left point of the shadow and the reference
\cite{AA}}
\end{figure}

For calculating these parameters, we consider five points
 $B=\left(\alpha_{t}, \beta_{t}\right)$, $D=\left(\alpha_{b}, \beta_{b}\right)$,
 $C=\left(\alpha_{r}, 0\right)$,
$A= \left(\alpha_{p}, 0\right)$, and $F=\left(\bar{\alpha}_{p},
0\right)$, which are top, bottom, rightmost, leftmost of the
shadow and leftmost of the reference circle respectively, so we
have
\begin{eqnarray}
R_{s}&=&\frac{\left(\alpha_{t}-\alpha_{r}\right)^{2}+\beta_{t}^{2}}{2\left(\alpha_{t}-\alpha_{r}\right)}\
,\nonumber\\
\delta_{s}&=&\frac{\left(\bar{\alpha}_{p}-\alpha_{p}\right)}{R_{s}}.
\end{eqnarray}

\begin{figure}[H]
\centering
\begin{subfigure}{.5\textwidth}
  \centering
  \includegraphics[width=.7\linewidth]{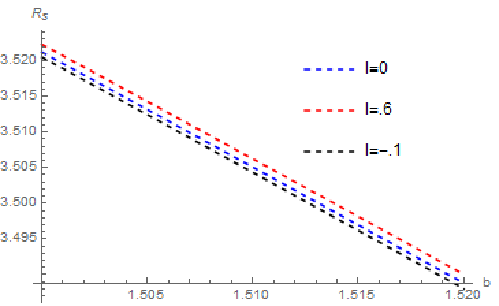}
\end{subfigure}%
\begin{subfigure}{.5\textwidth}
  \centering
  \includegraphics[width=.7\linewidth]{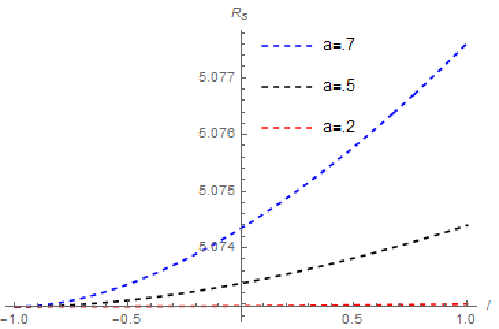}
\end{subfigure}
\caption{The left one gives variation of $R_{s}$ for various
values of $l$ with $a=.2$,  and $\theta=\pi/2$ and the right one
gives variation of $R_{s}$ against $l$ for various values of $a$
with $b=0.16$ and $\theta=\pi/2$ }
\begin{subfigure}{.5\textwidth}
\centering
  \includegraphics[width=.7\linewidth]{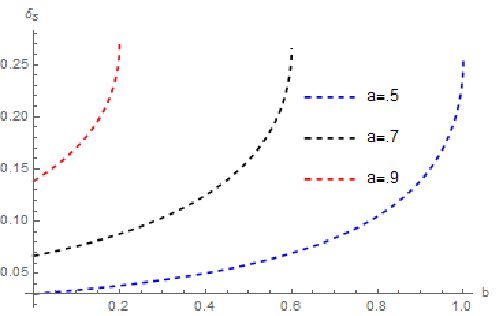}
\end{subfigure}%
\begin{subfigure}{.5\textwidth}
  \centering
  \includegraphics[width=.7\linewidth]{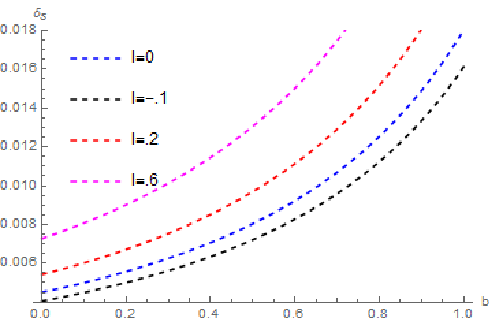}
\end{subfigure}
\caption{ The left one gives variation of $\delta_{s}$ for various
values of $a$ with  $l=0$ and $\theta=\pi/2$ and the right one
gives  variation of $\delta_{s}$ for various values of $l$ with
$a=0.2$ and $\theta=\pi/2$ }
\end{figure}

\begin{figure}[H]
\centering
\begin{subfigure}{.5\textwidth}
  \centering
  \includegraphics[width=.7\linewidth]{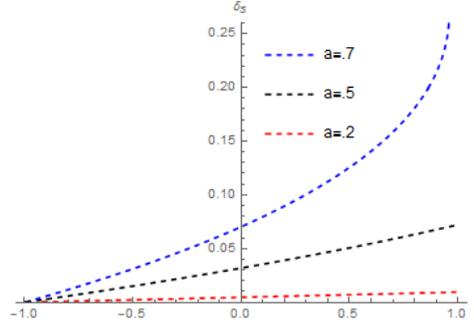}
\end{subfigure}
\caption{  The variation of $\delta_{s}$ against $l$ for various
values of a with $b=1$ and $\theta=\pi/2$ }
\end{figure}
From the plots show that the increase in $b$ for a fixed value $l$
and $a$ , leads to a decrease in $R_{s}$, and increases in
$\delta_{s}$. For fixed values of $b$, and $a$ both $R_{s}$ and
$\delta_{s}$ increases but the nature of variation differs for
different values of $a$. Thus the Lorentz Violating term has a
significant impact on the size of the shadow and its deviation
from the circular form. For all the plots, we have taken $M=1$ for
convenience.
\section{ENERGY EMISSION RATE}
We proceed in this Sec. with study the possible visibility of the
Kerr-Sen like black-hole through shadow. In the vicinity of
limiting constant value, the absorption cross-section of the
black-hole moderates lightly at high energy. We know that a
rotating black-hole can absorb electromagnetic waves, so the
absorbing cross-section \cite{BM} for a spherically symmetric
black-hole is.
\begin{equation}
\sigma_{l i m}=\pi R_{s}^{2}. \label{XSEC}
\end{equation}
Using the equation (\ref{XSEC}) we find out the energy emission
rate following the article \cite{AA}
\begin{equation}
\frac{d^{2} E}{d \omega d t}=\frac{2 \pi R_{s}^{2}}{e^{\left(\frac{\omega}{T}\right)}-1} \omega^{3}
\end{equation}
where $T$ is the Hawking temperature, $\omega$ the frequency of
radiation and $E$ stands for energy.
\begin{figure}[H]
\centering
\begin{subfigure}{.5\textwidth}
  \centering
  \includegraphics[width=.7\linewidth]{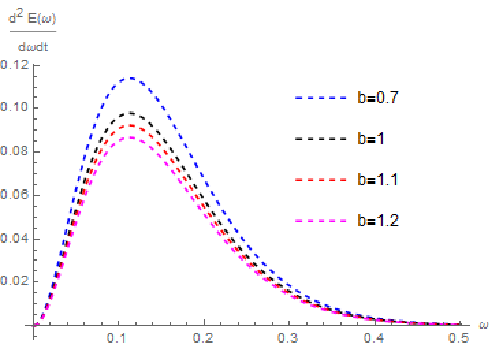}
\end{subfigure}%
\begin{subfigure}{.5\textwidth}
  \centering
  \includegraphics[width=.7\linewidth]{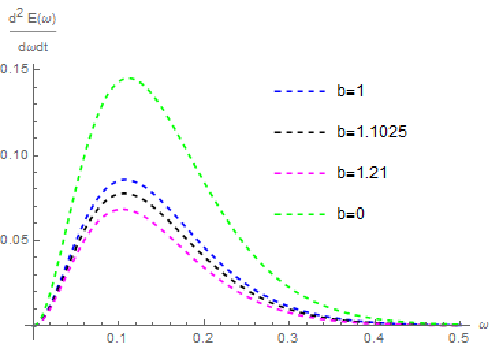}
\end{subfigure}
\caption{The left one gives variation of emission rate against
$\omega$ for various values of $b$ with $a=0$ and $l=0$ and the
right one gives variation of emission rate against $\omega$ for
various values of $b$ with $a=0.2$ and $l=0$}
\end{figure}

\begin{figure}[H]
\centering
\begin{subfigure}{.5\textwidth}
\centering
  \includegraphics[width=.7\linewidth]{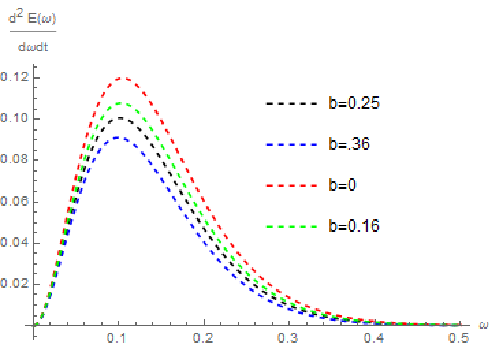}
\end{subfigure}%
\begin{subfigure}{.5\textwidth}
  \centering
  \includegraphics[width=.7\linewidth]{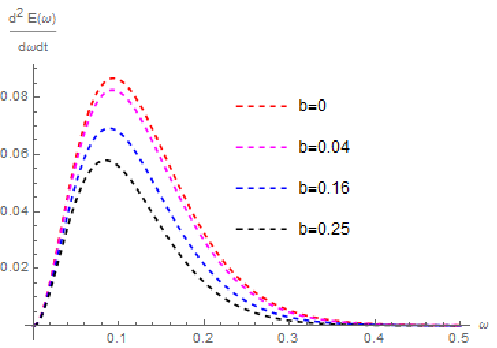}
\end{subfigure}
\caption{The left one gives variation of emission rate against
$\omega$ for various values of $b$ with $a=0.5$ and $l=0$ and the
 right one gives variation of emission rate against $\omega$ for various
 values of $b$ with $a=0.7$ and $l=0$}
\end{figure}

\begin{figure}[H]
\centering
\begin{subfigure}{.5\textwidth}
  \centering
  \includegraphics[width=.7\linewidth]{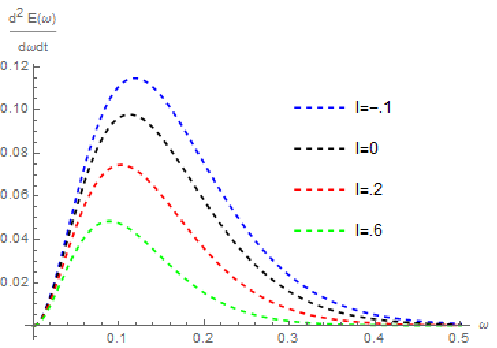}
\end{subfigure}%
\begin{subfigure}{.5\textwidth}
  \centering
  \includegraphics[width=.7\linewidth]{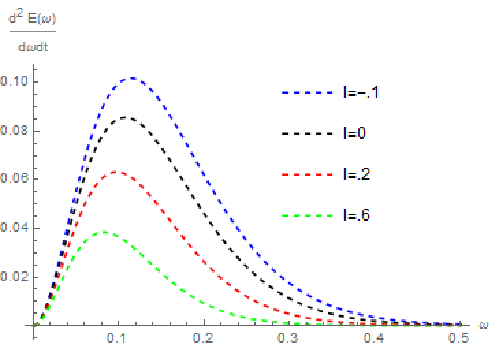}
\end{subfigure}
\caption{The left one gives variation of emission rate against
 $\omega$ for various values of $l$ with $a=0$ and $b=1$ and the
 right one gives variation of emission rate against $\omega$ for
 various values of $l$ with $a=0.2$ and $b=1$}
\begin{subfigure}{.5\textwidth}
\centering
  \includegraphics[width=.7\linewidth]{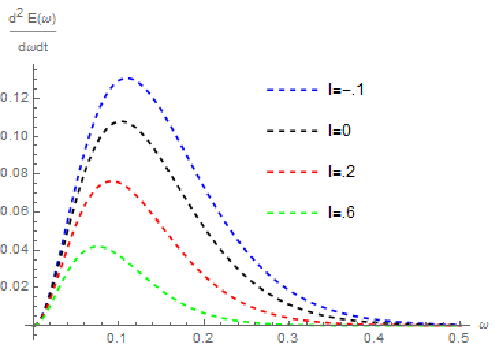}
\end{subfigure}%
\begin{subfigure}{.5\textwidth}
  \centering
  \includegraphics[width=.7\linewidth]{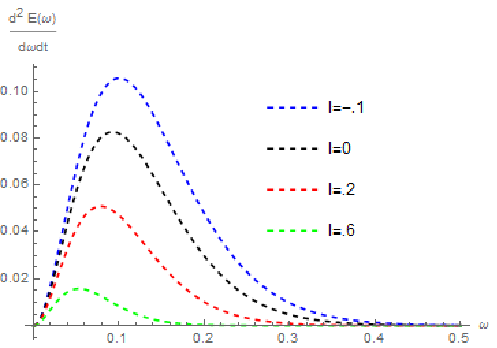}
\end{subfigure}
\caption{The left one gives variation of emission rate against $\omega$
for various values of $l$ with $a=0.5$ and $b=.16$ and the right gives variation
of emission rate against $\omega$ for various values of $l$ with $a=0.7$ and $b=.04$}
\end{figure}
We have sketched the energy emission rate versus $\omega$ for
various cases. It is clear from the sketch that the emission rate
decreases with an increase in $b$ for any set of fixed values of
$a$ and $\ell$. It also decreases when $\ell$ increases for any
set of fixed values of $a$ and $b$. However, there is a crucial
difference in the situation when $\ell$ increases. It is true that
emission rate decreases with the increase in $\ell$ like the case
when $b$ increases but unlike the situation when $b$ increases the
pick of the curve gets shifted when $\ell$ increases.

\section{Summary and discussion}
We have considered the Einstein-bumblebee gravity model where LV
scenario gets involved through a bumblebee field vector field
$B_\mu$. A spontaneous symmetry breaking allows the field to
acquires a vacuum expectation value that generates LV into the
system. A Kerr-Sen-like solution has been found out starting from
the generalized form of a radiating stationery axially symmetric
black-hole metric. For the parameter $b=0$ the matric turns into
Kerr-like metric and for both $b=0$ and $a=0$ the metric lands
onto Schwarzschild-like metric. The effective potential that
results from the null geodesics in the bumblebee rotating
black-hole spacetime is computed that in turn helps to get the
tentative nature of the motion of photon is predicted for
different choices of the parameters $a$, $b$ and $\ell$. The
nature of shadow is studied and how does the shadow gets deformed
that has also been studied for different variations of $a$, $b$,
and $\ell$. We observe that shadow gets shifted towards the right
for positive $\ell$ and shifted towards left for negative $\ell$
when $a$ and $b$ remains fixed. However, shifting is always
towards the right when $bl$ increases whatever the set of values
of $\ell$ and $a$ are taken. We have also studied the rate of
emission of energy for this type of black-hole. The emission rate
decreases when $b$ increases for any set of fixed values of $a$
and $\ell$. It also decreases when $\ell$ increases for an
arbitrary set of fixed values of $a$ and $b$. A crucial difference
however in the situation is noticed when $\ell$ increases. The
emission rate although decrease with the increase of $\ell$ like
the case when $b$ increases, the pick of the curve gets shifted
towards lower $\omega$ in the situation when $\ell$ increases. The
deformation of the shadow for Kerr-Sen-like black-hole in the
Einstein-bumblebee gravity model with the variation of $\ell$
observed here is a theoretical prediction. It shows that it
enhances the distortion of shadow, and it would be detected by the
new generation gravitational antennas.

\end{document}